\documentclass[a4paper,11pt]{article}
\usepackage{jheppub}

\usepackage{amsmath}
\usepackage{amssymb}
\usepackage{amsfonts}
\usepackage{graphicx}
\usepackage[T1]{fontenc}
\usepackage{slashed}
\usepackage{tikz}
\usetikzlibrary{arrows.meta,decorations.markings}
\usepackage{pgfplots}
\pgfplotsset{compat=1.16}
\usepackage{needspace}
\usepackage{microtype}
\usepackage{placeins}
\usepackage{booktabs}
\usepackage{dcolumn}

\hypersetup{
  pdftitle={Inclusive weak decays of the J/psi: single-charm decay and weak annihilation},
  pdfauthor={Lan-Bin Li, Chia-Wei Liu and Peng-Yuan Zheng},
  pdfsubject={Heavy-quark effective theory and vacuum saturation for inclusive weak charmonium decays}
}

\allowdisplaybreaks[2]

\newcommand{\jpsi}{J/\psi}
\newcommand{\Trc}{\operatorname{Tr}_{c}}
\newcommand{\Tweak}{\mathcal T}

\title{Inclusive weak decays of the $J/\psi$:\\
single-charm decay and weak annihilation}
\author{Lan-Bin Li,}
\emailAdd{lilanbin25@mails.ucas.ac.cn}
\author{Chia-Wei Liu}
\emailAdd{chiaweiliu@ucas.ac.cn}
\author{and Peng-Yuan Zheng}
\emailAdd{zhengpengyuan25@mails.ucas.ac.cn}
\affiliation{Hangzhou Institute for Advanced Study, University of Chinese
Academy of Sciences,\\ Hangzhou 310024, China}

\abstract{
Conventional estimates of $J/\psi$ weak decays, based on independent
single-charm decays, suggest a branching fraction of order $10^{-8}$.
Searches with samples of approximately $10^{10}$ $J/\psi$ events have
so far yielded upper limits on exclusive modes. We calculate the inclusive
weak branching fraction using the optical theorem, including single-charm
decay, charged- and neutral-current annihilation, and their interference.
The heavy-quark expansion identifies the leading
pair-annihilation structure and motivates vacuum saturation for the leading
color-singlet component. With tree-level
hard kernels and leading-logarithmic Wilson coefficients at
$\mu_0=M_{J/\psi}$, we obtain
$\mathcal B_{\rm weak}=(7.96^{+3.43}_{-0.43}\pm1.24)\times10^{-8}$.
The contribution mediated by the $W$ boson alone, i.e.\ single-charm
decays and $W$-exchange annihilation without the $W$--$Z$
interference, has a branching fraction of
$(2.37^{+1.64}_{-0.42}\pm0.13)\times10^{-8}$.
For the flavor sectors $(\Delta C,|\Delta S|)=(0,1)$ and
$|\Delta C|=1$, the branching fractions are
$(4.97^{+16.44}_{-4.97}\pm0.99)\times10^{-10}$ and
$(1.86^{+0.20}_{-0.06}\pm0.02)\times10^{-8}$, respectively,
including charge-conjugate channels. The first uncertainty estimates missing
perturbative corrections through scale variation over
$1~\mathrm{GeV}\leq\mu\leq3M_{J/\psi}$. The second is a hadronic   uncertainty, obtained by jointly varying
each vacuum-saturation and kinetic/chromomagnetic input within relative
ranges of
$\Lambda_{\rm QCD}/m_c\simeq20\%$. Pair annihilation dominates, with a substantial
neutral-current contribution and scale-sensitive $W$--$Z$ interference. 
}


\begin{document}
\maketitle

\section{Introduction}
\label{sec:introduction}

The $J/\psi$ allows weak interactions to be studied in a heavy
quark--antiquark bound state dominated by strong
and electromagnetic decays. Weak decay proceeds through the decay of either
charm constituent or $c\bar c$ annihilation, and the very small expected rates have motivated both theoretical
estimates and dedicated searches~\cite{Sanchis:1994,Sharma:1999,BES:2008weak,BESIII:2014weak}.
A widely used estimate treats the charm and anticharm as
independently decaying constituents, with a charmed-meson lifetime as
a proxy for the charm-quark decay time, so that the charm-changing width
is~\cite{Wang:2008nonleptonic}
\begin{equation}
 \Gamma_{\lvert\Delta C\rvert=1}\simeq 2\Gamma_D .
 \label{eq:intro-spectator-estimate}
\end{equation}
The factor of two accounts for the two constituents; with
$\Gamma_D=\Gamma_{D^+}$, the branching fraction is
$2\Gamma_D/\Gamma_{J/\psi}^{\rm tot}\simeq1.4\times10^{-8}$.

The applicability of the heavy-quark expansion (HQE) at the charm
scale has been examined in detail, including radiative and subleading
power corrections~\cite{Lenz:2015,Lenz:2013,King:2021}. For the $D^+$
meson, destructive Pauli interference and cancellations between
dimension-six and dimension-seven contributions make the absolute width
so sensitive to higher-order corrections and four-quark matrix elements
that, for several inputs, the truncated width has an unphysical negative
central value~\cite{King:2021,Friday:2025}. Agreement for lifetime ratios
does not by itself establish the accuracy of absolute charm widths; a
reorganization retaining phase-space-enhanced four-quark contributions in
the leading term has also been proposed~\cite{Mannel:2021}.

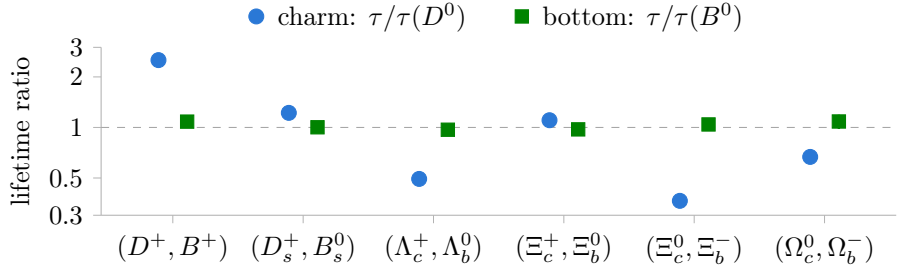
\begin{figure}[!htbp]
  \centering
  \begin{tikzpicture}
  \begin{axis}[
    width=0.80\textwidth, height=3.8cm,
    ymode=log, ymin=0.3, ymax=3, yminorticks=false,
    ytick={0.3,0.5,1,2,3}, yticklabels={0.3,0.5,1,2,3},
    ylabel={lifetime ratio},
    xmin=0.45, xmax=6.55, xtick={1,2,3,4,5,6},
    xticklabels={{$(D^+,B^+)$},{$(D_s^+,B_s^0)$},
      {$(\Lambda_c^+,\Lambda_b^0)$},{$(\Xi_c^+,\Xi_b^0)$},
      {$(\Xi_c^0,\Xi_b^-)$},{$(\Omega_c^0,\Omega_b^-)$}},
    axis lines=left, axis line style={-, gray!60}, tick style={gray!60},
    tick align=outside, major tick length=3pt,
    tick label style={font=\small}, label style={font=\small},
    legend style={at={(0.5,1.04)}, anchor=south, draw=none, fill=none,
      legend columns=2, font=\small,
      /tikz/every odd column/.append style={column sep=0.3em},
      /tikz/every even column/.append style={column sep=1.5em}},
    legend cell align=left,
  ]
  \addplot[forget plot, gray!70, dashed, domain=0.45:6.55, samples=2] {1};
  \addplot[only marks, mark=*, mark size=2.8pt, color={rgb,255:red,42;green,120;blue,214},
    error bars/.cd, y dir=both, y explicit, error mark=none,
    error bar style={line width=0.9pt}]
    coordinates {(0.89,2.5177) +- (0,0.0136) (1.89,1.2215) +- (0,0.0061) (2.89,0.4938) +- (0,0.0027) (3.89,1.1041) +- (0,0.0125) (4.89,0.3656) +- (0,0.0042) (5.89,0.6678) +- (0,0.0244)};
  \addlegendentry{charm: $\tau/\tau(D^0)$}
  \addplot[only marks, mark=square*, mark size=2.5pt, color={rgb,255:red,0;green,131;blue,0},
    error bars/.cd, y dir=both, y explicit, error mark=none,
    error bar style={line width=0.9pt}]
    coordinates {(1.11,1.0832) +- (0,0.0033) (2.11,1.0024) +- (0,0.0044) (3.11,0.9694) +- (0,0.0062) (4.11,0.9740) +- (0,0.0107) (5.11,1.0421) +- (0,0.0140) (6.11,1.0852) +- (0,0.1059)};
  \addlegendentry{bottom: $\tau/\tau(B^0)$}
  \end{axis}
  \end{tikzpicture}
  \caption{Singly heavy hadron lifetimes normalized to $\tau(D^0)$
  (circles) and $\tau(B^0)$ (squares)~\cite{PDG:2026}; paired baryons
  share their light-quark content.}
  \label{fig:lifetime-ratios}
\end{figure}

Lifetime differences among hadrons with the same heavy quark
arise mainly from four-quark spectator contributions, suppressed by
$(\Lambda_{\rm QCD}/m_Q)^3$ but enhanced by a phase-space factor of
order $16\pi^2$~\cite{Neubert:1997,Lenz:2015}. For bottom hadrons,
$\tau(B^+)/\tau(B^0)=1.083\pm0.003$ and
$\tau(\Lambda_b^0)/\tau(B^0)=0.970\pm0.006$~\cite{PDG:2026,LHCb:2014Lb},
in agreement with HQE predictions~\cite{Gratrex:2023bottom,Albrecht:2024};
for charmed hadrons, $\tau(D^+)/\tau(D^0)\simeq2.52$ and
$\tau(\Lambda_c^+)/\tau(D^0)\simeq0.49$~\cite{PDG:2026}.
Figure~\ref{fig:lifetime-ratios} shows that charm lifetimes span a factor
of about seven, whereas bottom ones vary by only about $10\%$.

The $\Omega_c^0$ is a striking example: long regarded as the
shortest-lived singly charmed baryon, it was found to live nearly four
times longer than the previous world
average~\cite{LHCb:2018Omc,LHCb:2022promptOmc,BelleII:2023Omc}. It has
been argued that the HQE for the $\Omega_c^0$ breaks down at order
$1/m_c^4$ and that a consistent expansion requires
$\tau(\Omega_c^0)>\tau(\Lambda_c^+)$~\cite{Cheng:2022Omc}, a hierarchy
accommodated by later analyses~\cite{Gratrex:2022,Cheng:2023singly}.
Doubly charmed baryons show a similarly large splitting, driven by Pauli
interference and $W$
exchange~\cite{Kiselev:1999,Guberina:1999,Cheng:2018doubly,Dulibic:2023,Cheng:2026doubly,Dulibic:2026,LHCb:2018Xicc}.

Exclusive weak channels of the $J/\psi$ have been studied using
heavy-hadron spin symmetry~\cite{Sanchis:1994}, quark-model form factors
combined with factorization~\cite{Sharma:1999}, QCD sum
rules~\cite{Wang:2008semileptonic,Wang:2008nonleptonic}, covariant
light-front form factors~\cite{Shen:2008}, QCD factorization with
radiative corrections~\cite{Sun:2015QCDF}, and a perturbative-QCD
approach~\cite{Sun:2016PQCD}. These approaches relate individual decay
rates to $J/\psi\to D_{(s)}^{(*)}$ transition form factors and, for
nonleptonic modes, rely on assumptions about hadronic factorization. The
summed branching fraction of the nonleptonic two-body spectator channels
can be as large as $1.3\times10^{-8}$~\cite{Wang:2008nonleptonic}.
Such calculations provide channel-specific targets for experiment, but
they address only charm-changing final states. Pair annihilation, which
produces final states without open charm, requires an inclusive treatment
in which single-charm decay and annihilation are organized together.

Substantial experimental effort has been devoted to these rare
decays. Searches progressed from $5.77\times10^7$ $J/\psi$ events at
BESII~\cite{BES:2008weak} to $2.25\times10^8$ events in early BESIII
studies~\cite{BESIII:2014weak}, and subsequently to about $10^{10}$
events~\cite{BESIII:2021Denu,BESIII:2025Dshad,BESIII:2026Dsenu}. With
these samples, BESIII obtained the 90\% confidence-level limits
$\mathcal B(J/\psi\to D^-e^+\nu_e+\mathrm{c.c.})<7.1\times10^{-8}$ and
$\mathcal B(J/\psi\to D_s^-e^+\nu_e+\mathrm{c.c.})<9.9\times10^{-8}$~\cite{BESIII:2021Denu,BESIII:2026Dsenu},
and limits of $8.0\times10^{-7}$ and $4.1\times10^{-7}$ on
$D_s^-\rho^+$ and $D_s^-\pi^+$, including charge-conjugate
modes~\cite{BESIII:2025Dshad}. The program also covers further hadronic
modes containing a $D$ meson~\cite{BESIII:2024Dhad}, the semimuonic
channel $J/\psi\to D^-\mu^+\nu_\mu+\mathrm{c.c.}$~\cite{BESIII:2024Dmunu},
and the flavor-changing neutral-current mode
$J/\psi\to D^0e^+e^-+\mathrm{c.c.}$~\cite{BESIII:2017FCNC}. No
significant signal has been observed. A proposed super tau-charm facility
would collect data samples about two orders of magnitude
larger~\cite{STCF:2024}, bringing branching fractions of order
$10^{-8}$ within reach.

Sections~\ref{sec:weak-topologies-rg}--\ref{sec:numerical-analysis-rg}
construct the transition operator, develop the heavy-pair expansion,
evaluate the matrix elements, and present the numerical results;
section~\ref{sec:conclusions} concludes.

\section{Theoretical framework}
\label{sec:weak-topologies-rg}

The weak Hamiltonian retained here is
\begin{equation}
 \mathcal H_{\rm eff}
 =\bigl[\mathcal H_{\rm NL}^{c}
          +\mathcal H_{\rm SL}^{c}+\mathrm{h.c.}\bigr]
   +\mathcal H_{\rm NL}+\mathcal H_{Z,q}+\mathcal H_{Z,\ell} .
 \label{eq:weak-hamiltonian-rg}
\end{equation}
Here NL and SL denote nonleptonic and semileptonic decays, and h.c.\ denotes
the Hermitian-conjugate Hamiltonians describing the corresponding
single-$\bar c$ decays.  We use
$\Gamma_L^\mu=\gamma^\mu(1-\gamma_5)$ and
$\Gamma_R^\mu=\gamma^\mu(1+\gamma_5)$.
The charged-current terms, with sums over $q_1,q_2=d,s$ and
$\ell=e,\mu$, take the form~\cite{Buras:1990}:
\begin{equation}
 \mathcal H_{\rm NL}^c 
 =\frac{G_F}{\sqrt2}\sum_{q_1,q_2}
 V_{cq_1}^*V_{uq_2}
 \Bigl[
 C_1(\bar q_{1i}\Gamma_L^\mu c^i)
 (\bar u_j\Gamma_{L\mu}q_2^j)
 +C_2(\bar q_{1i}\Gamma_L^\mu c^j)
 (\bar u_j\Gamma_{L\mu}q_2^i)
 \Bigr] .
 \label{eq:HNL-rg}
\end{equation}
\begin{equation}
 \mathcal H_{\rm SL}^{c}
 =\frac{G_F}{\sqrt2}
   \sum_{q_1,\ell}V_{cq_1}^*
   (\bar q_{1i}\Gamma_L^\mu c^i)
   (\bar\nu_\ell\Gamma_{L\mu}\ell) .
 \label{eq:HSL-rg}
\end{equation}
\begin{equation}
 \mathcal H_{\rm NL} 
 =\frac{G_F}{\sqrt2}\sum_{q_1,q_2}
 V_{cq_1}^*V_{cq_2}
 \Bigl[
 C_1(\bar q_{1i}\Gamma_L^\mu c^i)
 (\bar c_j\Gamma_{L\mu}q_2^{j})
 +C_2(\bar q_{1i}\Gamma_L^\mu c^j)
 (\bar c_j\Gamma_{L\mu}q_2^{i})
 \Bigr] .
 \label{eq:HW-rg}
\end{equation}
Here $G_F$ is the Fermi constant, $V$ is the CKM matrix, and
$C_{1,2}$ are Wilson coefficients. 
The superscript $c$ marks the charm-number-violating terms, while
$\mathcal H_{\rm NL}$ conserves charm and induces $W$ exchange. 
 The neutral-current terms take the form~\cite{Du:2024nc}:
\begin{equation}
 \mathcal H_{Z,q}
 =\sqrt2\,G_F\sum_{q,\chi,\chi'}
   g_{Z,\chi}^{c}g_{Z,\chi'}^{q}
   \Bigl[
   C_Z^{\chi\chi'}(\bar c_i\Gamma_\chi^\mu c^i)(\bar q_j\Gamma_{\chi'\mu}q^j)
   +\widetilde C_Z^{\chi\chi'}(\bar c_i\Gamma_\chi^\mu c^j)(\bar q_j\Gamma_{\chi'\mu}q^i)
   \Bigr] .
 \label{eq:HZ-quark-rg}
\end{equation}
\begin{equation}
 \mathcal H_{Z,\ell}
 =\sqrt2\,G_F
   \sum_{f,\chi,\chi'}
   g_{Z,\chi}^{c}g_{Z,\chi'}^{f}
   (\bar c_i\Gamma_\chi^\mu c^i)
   (\bar f\Gamma_{\chi'\mu}f) .
 \label{eq:HZ-lepton-rg}
\end{equation}
The sums run over $q=u,d,s$, $f=e,\mu,\nu_e,\nu_\mu,\nu_\tau$, and
$\chi,\chi'=L,R$.  Here $i,j$ are color indices, with $N_c=3$ colors.  The neutral-current Wilson-coefficient combinations appearing in
eq.~\eqref{eq:HZ-quark-rg} are expressed in the $C_i^Z$ basis
of~\cite{Du:2024nc} as
\begin{equation}
\begin{aligned}
 C_Z^{LL}
 &=C_Z^{RR}=\frac{C_1^Z+C_2^Z}{2},
 &\qquad
 \widetilde C_Z^{LL}
 &=\widetilde C_Z^{RR}=\frac{C_1^Z-C_2^Z}{2},
  \\
 C_Z^{LR}
 &=C_Z^{RL}=C_4^Z,
 &
 \widetilde C_Z^{LR}
 &=\widetilde C_Z^{RL}=\frac{C_3^Z-C_4^Z}{N_c} .
\end{aligned}
\label{eq:CZ-combinations-rg}
\end{equation}
Here $C_i^Z$, $i=1,\dots,4$, are the neutral-current Wilson
coefficients at the scale $\mu$, normalized to $C_i^Z(M_Z)=1$.
The chiral neutral-current couplings are
\begin{equation}
 g_{Z,L}
 =T_3-Qs_W^2,
 \qquad
 g_{Z,R}
 =-Qs_W^2 .
 \label{eq:chiral-Z-couplings-rg}
\end{equation}
Here $s_W=\sin\theta_W$, $Q$ is the electric charge of the fermion, and $T_3$ is its
weak-isospin eigenvalue.
For real Wilson coefficients, the ordered flavor sum makes
$\mathcal H_{\rm NL}$ Hermitian, while
$\mathcal H_Z\equiv\mathcal H_{Z,q}+\mathcal H_{Z,\ell}$ is Hermitian term by
term.  The
$Z$ field has been integrated out, so $\mathcal H_Z$ is represented by a
local four-fermion vertex and
no separate propagator factor is retained~\cite{Du:2024nc}.

Throughout this work, $\Gamma_{\rm weak}$ denotes the weak-only contribution
at order $G_F^2$.  With the effective Hamiltonian specified above, the
optical theorem expresses this inclusive weak decay width through the
absorptive part of the forward-scattering amplitude with two Hamiltonian
insertions~\cite{Bigi:1992,Manohar:1994}:
\begin{equation}
 \begin{aligned}
 \Gamma_{\rm weak}(\jpsi)
 &=\frac{1}{2M_{J/\psi}}\operatorname{Im}\,i\!\int d^4x\,
   \langle\jpsi|\mathrm T\!\left\{
   \mathcal H_{\rm eff}(x)\mathcal H_{\rm eff}(0)
   \right\}|\jpsi\rangle
 \\
 &=\frac{1}{2M_{J/\psi}}
   \langle\jpsi|\Tweak|\jpsi\rangle .
 \end{aligned}
 \label{eq:optical-width-rg}
\end{equation}
Here $M_{J/\psi}$ is the $\jpsi$ mass, and $\Tweak$ denotes the inclusive
transition operator generated by the double Hamiltonian insertion.

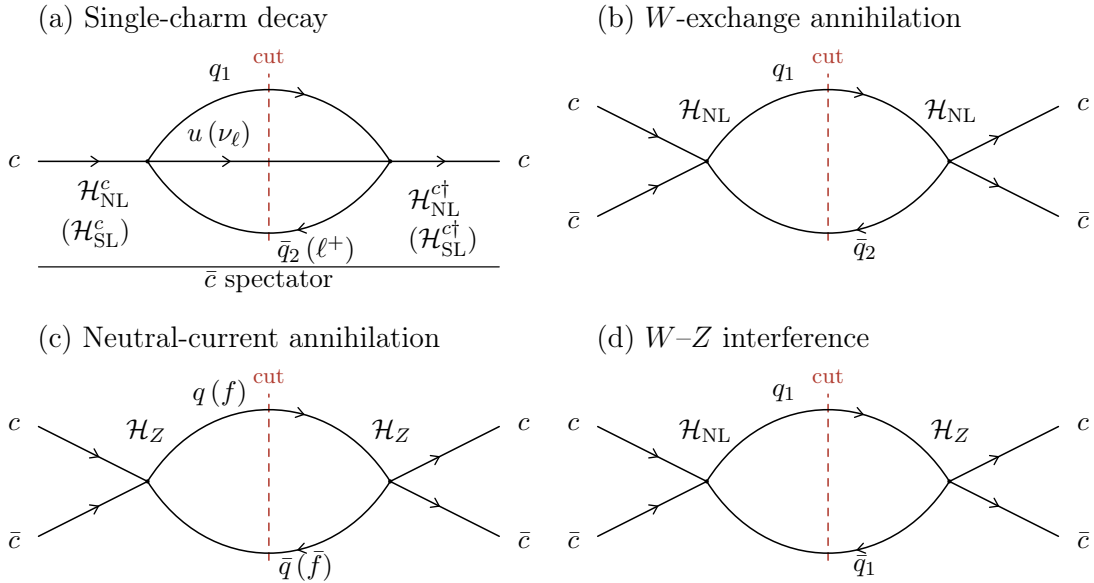
\begin{figure}[!htbp]
  \centering
  \resizebox{\linewidth}{!}{
\definecolor{cutred}{HTML}{B03A2E}
\tikzset{
  fw/.style={line width=0.62pt},
  halfarrow/.style={postaction={decorate,decoration={markings,
      mark=at position #1 with {%
        \draw[line width=0.55pt] (-0.105,0.072)--(0,0)--(-0.105,-0.072);}}}},
  halfarrow/.default=0.6,
  dot/.style={circle,fill=black,inner sep=0pt,minimum size=1.7pt},
  cutline/.style={cutred,dashed,line width=0.62pt},
}

\newcommand{\wloop}{%
  \draw[fw,halfarrow=0.62] (0,0) .. controls (0.86,1.36) and (2.59,1.36) .. (3.45,0);
  \draw[fw,halfarrow=0.40] (3.45,0) .. controls (2.59,-1.36) and (0.86,-1.36) .. (0,0);
  \node[dot] at (0,0) {};
  \node[dot] at (3.45,0) {};}

\newcommand{\wcut}{%
  \draw[cutline] (1.725,-1.12) -- (1.725,1.24);
  \node[cutred,font=\footnotesize,anchor=south] at (1.725,1.27) {cut};}

\begin{tikzpicture}[font=\normalsize,line cap=round]

  \begin{scope}[shift={(0,0)}]
    \node[anchor=west,font=\large] at (-1.70,2.02) {(a) Single-charm decay};
    \wcut
    \draw[fw,halfarrow=0.55] (-1.55,0) -- (0,0);
    \draw[fw,halfarrow=0.50] (3.45,0) -- (5.00,0);
    \draw[fw,halfarrow=0.34] (0,0) -- (3.45,0);
    \wloop
    \node[anchor=east] at (-1.68,0) {$c$};
    \node[anchor=west] at (5.13,0) {$c$};
    \node at (1.04,1.26) {$q_1$};
    \node at (1.02,0.38) {$u\,(\nu_\ell)$};
    \node at (2.40,-1.26) {$\bar q_2\,(\ell^+)$};
    \node[anchor=north east,align=right] at (-0.11,-0.14)
      {$\mathcal H^c_{\mathrm{NL}}$\\[1pt]$(\mathcal H^c_{\mathrm{SL}})$};
    \node[anchor=north west,align=left] at (3.57,-0.14)
      {$\mathcal H^{c\dagger}_{\mathrm{NL}}$\\[1pt]$(\mathcal H^{c\dagger}_{\mathrm{SL}})$};
    \draw[line width=0.5pt] (-1.55,-1.50) -- (5.00,-1.50);
    \node at (1.725,-1.70) {$\bar c$ spectator};
  \end{scope}

  \begin{scope}[shift={(7.95,0)}]
    \node[anchor=west,font=\large] at (-1.70,2.02) {(b) $W$-exchange annihilation};
    \wcut
    \draw[fw,halfarrow=0.55] (-1.55,0.78) -- (0,0);
    \draw[fw,halfarrow=0.55] (-1.55,-0.78) -- (0,0);
    \draw[fw,halfarrow=0.45] (3.45,0) -- (5.00,0.78);
    \draw[fw,halfarrow=0.45] (3.45,0) -- (5.00,-0.78);
    \wloop
    \node[anchor=east] at (-1.68,0.82) {$c$};
    \node[anchor=east] at (-1.68,-0.84) {$\bar c$};
    \node[anchor=west] at (5.13,0.82) {$c$};
    \node[anchor=west] at (5.13,-0.84) {$\bar c$};
    \node at (1.10,1.26) {$q_1$};
    \node at (2.25,-1.26) {$\bar q_2$};
    \node at (-0.02,0.72) {$\mathcal H_{\mathrm{NL}}$};
    \node at (3.47,0.72) {$\mathcal H_{\mathrm{NL}}$};
  \end{scope}

  \begin{scope}[shift={(0,-4.55)}]
    \node[anchor=west,font=\large] at (-1.70,2.02) {(c) Neutral-current annihilation};
    \wcut
    \draw[fw,halfarrow=0.55] (-1.55,0.78) -- (0,0);
    \draw[fw,halfarrow=0.55] (-1.55,-0.78) -- (0,0);
    \draw[fw,halfarrow=0.45] (3.45,0) -- (5.00,0.78);
    \draw[fw,halfarrow=0.45] (3.45,0) -- (5.00,-0.78);
    \wloop
    \node[anchor=east] at (-1.68,0.82) {$c$};
    \node[anchor=east] at (-1.68,-0.84) {$\bar c$};
    \node[anchor=west] at (5.13,0.82) {$c$};
    \node[anchor=west] at (5.13,-0.84) {$\bar c$};
    \node at (1.03,1.26) {$q\,(f)$};
    \node at (2.25,-1.20) {$\bar q\,(\bar f)$};
    \node at (-0.02,0.72) {$\mathcal H_Z$};
    \node at (3.47,0.72) {$\mathcal H_Z$};
  \end{scope}

  \begin{scope}[shift={(7.95,-4.55)}]
    \node[anchor=west,font=\large] at (-1.70,2.02) {(d) $W$--$Z$ interference};
    \wcut
    \draw[fw,halfarrow=0.55] (-1.55,0.78) -- (0,0);
    \draw[fw,halfarrow=0.55] (-1.55,-0.78) -- (0,0);
    \draw[fw,halfarrow=0.45] (3.45,0) -- (5.00,0.78);
    \draw[fw,halfarrow=0.45] (3.45,0) -- (5.00,-0.78);
    \wloop
    \node[anchor=east] at (-1.68,0.82) {$c$};
    \node[anchor=east] at (-1.68,-0.84) {$\bar c$};
    \node[anchor=west] at (5.13,0.82) {$c$};
    \node[anchor=west] at (5.13,-0.84) {$\bar c$};
    \node at (1.10,1.26) {$q_1$};
    \node at (2.25,-1.20) {$\bar q_1$};
    \node at (-0.02,0.72) {$\mathcal H_{\mathrm{NL}}$};
    \node at (3.47,0.72) {$\mathcal H_Z$};
  \end{scope}

\end{tikzpicture}}
  \caption{Forward-scattering topologies entering the inclusive weak
  transition operator: (a) single-charm decay; (b) $W$-exchange annihilation;
  (c) neutral-current annihilation; and (d) $W$--$Z$ interference.
  Dashed lines
  mark unitarity cuts; dots represent local weak-Hamiltonian insertions.}
  \label{fig:weak-mechanisms-rg}
\end{figure}

Substituting the effective Hamiltonian into
eq.~\eqref{eq:optical-width-rg} and organizing the double insertions by the
topologies in figure~\ref{fig:weak-mechanisms-rg} gives
\begin{equation}\label{eq:topology-decomposition-rg}
 \Tweak
 =\mathcal T_{c+\bar c}
   +\mathcal T_{c\bar c},
 \qquad
 \mathcal T_{c\bar c}
 =\mathcal T_{\rm WE}
   +\mathcal T_{\rm NC}
   +\mathcal T_{\rm int}.
\end{equation}
The first term contains the decay contributions of the charm and anticharm
constituents.  Each term in $\mathcal T_{c\bar c}$ contains the
complete heavy pair: $\mathcal T_{\rm WE}$, $\mathcal T_{\rm NC}$, and
$\mathcal T_{\rm int}$ describe $W$-exchange annihilation, neutral-current
annihilation, and their interference.

After local matching, the transition operator is organized schematically
in the full-QCD basis as
\begin{equation}
 \Tweak
 =\frac{G_F^2m_c^5}{192\pi^3}
 \left[
 \left(
 c_3\mathcal O_3
 +\frac{c_5}{m_c^2}\mathcal O_5
 \right)_{2}
 +16\pi^2
 \sum_n\left(
 \frac{\mathcal C_n^{(8)}}{m_c^5}\mathcal Q_n
 \right)_{4}
 +\cdots
 \right] .
 \label{eq:transition-operator-ope-rg}
\end{equation}
The first term in eq.~\eqref{eq:transition-operator-ope-rg} describes the
single-charm decay sector through the order retained here, while the second
describes the leading heavy-pair annihilation kernels.  In the latter,
$\mathcal Q_n$ denotes a full-QCD four-charm operator with two current
derivatives and canonical dimension eight.  The index $n$ labels the
topology, chirality, and color contraction, and
$\mathcal C_n^{(8)}$ is a dimensionless short-distance coefficient. The
subscripts $2$ and $4$ mark two- and four-quark operators.

The leading two-quark term arises from the three-body cut of the partonic
$1\to3$ single-charm decay, whereas the four-quark terms arise from the
two-body cuts of the $2\to2$ heavy-pair annihilation processes.  Relative to
the common three-body normalization in
eq.~\eqref{eq:transition-operator-ope-rg}, the latter carry the conventional
phase-space enhancement $16\pi^2$~\cite{Neubert:1997,Lenz:2013}. The remaining
numerical and mass-dependent factors are contained in the short-distance
coefficients.

For figure~\ref{fig:weak-mechanisms-rg}(a) and its charge-conjugate counterpart,
contracting the hard final-state fields gives the standard local
OPE~\cite{Bigi:1992,Blok:1993,Manohar:1994}. Using
the original relativistic charm field, the two-quark sector is
\begin{equation}
 \mathcal T_{c+\bar c}
 =\frac{G_F^2m_c^5}{192\pi^3}
 \left[
 c_3\mathcal O_3
 +\frac{c_5}{m_c^2}\mathcal O_5
 +\sum_{d,n}\frac{c_{d,n}}{m_c^{d-3}}\mathcal O_{d,n}
 \right] .
 \label{eq:constituent-decay-full-qcd-rg}
\end{equation}
The first two operators in this expansion are
\begin{equation}
 \mathcal O_3
 =\bar c c,
 \qquad
 \mathcal O_5
 =\bar c\,\frac{g_s}{2}\sigma_{\mu\nu}G^{\mu\nu}c .
 \label{eq:O3-O5-rg}
\end{equation}
Here $m_c$ is the charm-quark mass in the pole convention, and $g_s$ is the
QCD gauge coupling. 
Writing $\mathcal A_\mu=\mathcal A_\mu^aT^a$, we use
$D_\mu=\partial_\mu-ig_s\mathcal A_\mu$ and
$[D_\mu,D_\nu]=-ig_sG_{\mu\nu}$, where
$G_{\mu\nu}=G_{\mu\nu}^aT^a$.  The index $a$ is an adjoint color index,
and the fundamental generators obey
$\operatorname{Tr}(T^aT^b)=\delta^{ab}/2$.  Finally,
$\sigma_{\mu\nu}=i[\gamma_\mu,\gamma_\nu]/2$.  In the last term, $d$ runs over
the canonical dimensions $d\geq6$, and $n$ labels the independent
operators at fixed $d$.

\Needspace{5\baselineskip}
At leading order, neglecting light-fermion masses and summing over the
charged-current quark and charged-lepton flavors specified above, the two
coefficients are
\begin{equation}\label{eq:charm-decay-bilinear-coefficients-rg}
\begin{aligned}
 c_3(\mu)
 &=\bigl(|V_{cs}|^2+|V_{cd}|^2\bigr)
 \Bigl\{2+\bigl(|V_{ud}|^2+|V_{us}|^2\bigr)
 \bigl[N_c(C_1^2+C_2^2)+2C_1C_2\bigr]\Bigr\},
 \\
 c_5(\mu)
 &=-2\bigl(|V_{cs}|^2+|V_{cd}|^2\bigr)
 \Bigl\{2+\bigl(|V_{ud}|^2+|V_{us}|^2\bigr)
 \bigl[N_c(C_1^2+C_2^2)+10C_1C_2\bigr]\Bigr\}.
\end{aligned}
\end{equation}
For the heavy-pair topologies in
figures~\ref{fig:weak-mechanisms-rg}(b)--(d), the light fields are contracted
between the two Hamiltonian insertions while all four charm fields remain
uncontracted.  The corresponding forward-scattering contributions, with
$\mathcal T_{\rm NC}=\mathcal T_{{\rm NC},q}+\mathcal T_{{\rm NC},\ell}$, are
\begin{align}
 \mathcal T_{\rm WE}
 &=-\frac{G_F^2}{12\pi}
 \sum_{q_1,q_2}|V_{cq_1}|^2|V_{cq_2}|^2
 \left[
 C_1^2\widetilde{\mathcal Q}[L,L]
 +(N_cC_2^2+2C_1C_2)\mathcal Q[L,L]
 \right],
 \label{eq:T8-W-rg}
 \\
 \mathcal T_{{\rm NC},q}
 &=-\frac{G_F^2}{3\pi}
 \sum_{q,\chi,\chi',\chi''}(g_{Z,\chi''}^{q})^2
 g_{Z,\chi}^{c}g_{Z,\chi'}^{c}
 \left[
 \mathcal K_{ZZ}^{\chi\chi';\chi''}
 \mathcal Q[\chi,\chi']
 +\widetilde{\mathcal K}_{ZZ}^{\chi\chi';\chi''}
 \widetilde{\mathcal Q}[\chi,\chi']
 \right],
 \label{eq:T8-Z-quark-rg}
 \\
 \mathcal T_{{\rm NC},\ell}
 &=-\frac{G_F^2}{3\pi}
 \sum_{f,\chi,\chi',\chi''}(g_{Z,\chi''}^{f})^2
 g_{Z,\chi}^{c}g_{Z,\chi'}^{c}
 \mathcal Q[\chi,\chi'],
 \label{eq:T8-Z-lepton-rg}
 \\*
 \mathcal T_{\rm int}
 &=-\frac{G_F^2}{3\pi}
 \sum_{q_1,\chi}|V_{cq_1}|^2g_{Z,L}^{q_1}g_{Z,\chi}^{c}
 \left[
 \mathcal K_{WZ}^{\chi}
 \mathcal Q[L,\chi]
 +\widetilde{\mathcal K}_{WZ}^{\chi}
 \widetilde{\mathcal Q}[L,\chi]
 \right].
 \label{eq:T8-WZ-rg}
\end{align}
Gauge covariance requires the derivative of a color-open heavy-quark
bilinear to be
\begin{equation}
 \mathcal D_\mu(\bar c_j\Gamma_\chi^\nu c^i)
 =(\bar c\overleftarrow D_\mu)_j\Gamma_\chi^\nu c^i
  +\bar c_j\Gamma_\chi^\nu(D_\mu c)^i .
 \label{eq:current-covariant-derivative-rg}
\end{equation}
We write both color contractions using covariant derivatives.  For the
color-singlet bilinears in the untilded contraction, these derivatives
reduce to ordinary derivatives.  The two color contractions are
\begin{align}
 \mathcal Q[\chi,\chi']
 &=\mathcal D_\alpha(\bar c_i\Gamma_{\chi\rho}c^i)
 \mathcal D^\alpha(\bar c_j\Gamma_{\chi'}^\rho c^j)
 -\mathcal D_\rho(\bar c_i\Gamma_\chi^\rho c^i)
  \mathcal D_\sigma(\bar c_j\Gamma_{\chi'}^\sigma c^j),
 \label{eq:dimension-eight-operator-definitions-rg}\\*[0.3em]
 \widetilde{\mathcal Q}[\chi,\chi']
 &=\mathcal D_\alpha(\bar c_j\Gamma_{\chi\rho}c^i)
 \mathcal D^\alpha(\bar c_i\Gamma_{\chi'}^\rho c^j)
 -\mathcal D_\rho(\bar c_j\Gamma_\chi^\rho c^i)
  \mathcal D_\sigma(\bar c_i\Gamma_{\chi'}^\sigma c^j).
 \label{eq:dimension-eight-color-rearranged-rg}
\end{align}
Both $\mathcal Q[\chi,\chi']$ and
$\widetilde{\mathcal Q}[\chi,\chi']$ are symmetric under
$\chi\leftrightarrow \chi'$.  The three unordered chirality pairs
$LL,LR,RR$, together with the untilded and tilded color contractions,
therefore give six independent full-QCD operators.

The combinations $\mathcal K_{ZZ}$, $\widetilde{\mathcal K}_{ZZ}$,
$\mathcal K_{WZ}$, and $\widetilde{\mathcal K}_{WZ}$ entering the
neutral-current and interference terms in eqs.~\eqref{eq:T8-Z-quark-rg}--\eqref{eq:T8-WZ-rg}
are defined by
\begin{equation}
 \mathcal K_{ZZ}^{\chi\chi';\chi''}
 =N_c C_Z^{\chi\chi''}C_Z^{\chi'\chi''}
  +C_Z^{\chi\chi''}\widetilde C_Z^{\chi'\chi''}
  +\widetilde C_Z^{\chi\chi''}C_Z^{\chi'\chi''},
 \qquad
 \widetilde{\mathcal K}_{ZZ}^{\chi\chi';\chi''}
 =\widetilde C_Z^{\chi\chi''}\widetilde C_Z^{\chi'\chi''} .
 \label{eq:KZZ-rg}
\end{equation}
\begin{equation}
 \mathcal K_{WZ}^{\chi}
 =(C_1+N_cC_2)C_Z^{\chi L}+C_2\widetilde C_Z^{\chi L},
 \qquad
 \widetilde{\mathcal K}_{WZ}^{\chi}
 =C_1\widetilde C_Z^{\chi L} .
 \label{eq:KWZ-rg}
\end{equation}
All Wilson coefficients entering eqs.~\eqref{eq:KZZ-rg} and \eqref{eq:KWZ-rg} use the same scale
$\mu$.

\FloatBarrier
\section{Heavy-pair expansion through canonical dimension seven}
\label{sec:heavy-pair-expansion-rg}

The heavy-quark effective-field expansion identifies the leading
annihilation structure of the four-charm operators and motivates the
vacuum-saturation approximation used in
section~\ref{sec:hadronic-matrix-elements-rg}.
For the velocity-projector construction we use the standard heavy-field
expansion~\cite{Neubert:1994}. The heavy-pair power counting and color-singlet treatment
follow~\cite{Beneke:1996,BBL:1995}.
Let $v^\mu$ be the heavy-pair four-velocity, normalized by $v^2=1$.
The charm field is decomposed as
\begin{equation}
 c(x)
 =e^{-im_cv\cdot x}h_+(x)
  +e^{+im_cv\cdot x}h_-(x)
  +\mathcal O\!\left(\frac1{m_c}\right) .
 \label{eq:heavy-frequency-fields-rg}
\end{equation}
The corresponding projectors satisfy
\begin{equation}
 P_\pm h_\pm
 =h_\pm,
 \qquad
 P_\pm=\frac{1\pm\slashed v}{2} .
 \label{eq:heavy-projectors-rg}
\end{equation}
The two annihilation currents and their transverse projections are defined by
\begin{equation}
 \begin{aligned}
 (a_{\chi\mu}^{[0]})^i{}_j
 &=\bar h_{-,j}\Gamma_{\chi\mu}h_+^i,
 \\
 (a_{\chi\mu}^{[1]})^i{}_j
 &=\bar h_{-,j}
   \bigl(-i\overleftarrow{\slashed D}_{\!\perp}\bigr)
   \Gamma_{\chi\mu}h_+^i
   +\bar h_{-,j}\Gamma_{\chi\mu}
   i\slashed D_{\!\perp}h_+^i,
 \\
 D_\perp^\mu
 &=D^\mu-v^\mu v\!\cdot D,
 \qquad
 \mathcal D_\perp^\mu
 =\mathcal D^\mu-v^\mu v\!\cdot\mathcal D,
 \\
 a_{\chi\perp}^{[r]\mu}
 &=(g^{\mu\nu}-v^\mu v^\nu)a_{\chi\nu}^{[r]},
 \qquad r=0,1.
 \end{aligned}
 \label{eq:pair-annihilation-current-rg}
\end{equation}
Here $a_\chi^{[0]\mu}$ is the leading local pair-annihilation
current, while $a_\chi^{[1]\mu}$ contains one residual derivative.

The full current also contains number-conserving heavy-field bilinears.
At leading order its frequency decomposition is
\begin{equation}
\begin{aligned}
\bar c_j\Gamma_\chi^\mu c^i
&=(n_\chi^{[0]\mu})^i{}_{j}
 +e^{-2im_cv\cdot x}(a_\chi^{[0]\mu})^i{}_{j}
 +e^{+2im_cv\cdot x}[(a_\chi^{[0]\mu})^\dagger]^i{}_{j}
 +\mathcal O\!\left(\frac1{m_c}\right),\\
(n_\chi^{[0]\mu})^i{}_{j}
&=\bar h_{+,j}\Gamma_\chi^\mu h_+^i
 +\bar h_{-,j}\Gamma_\chi^\mu h_-^i.
\end{aligned}
\label{eq:heavy-current-frequency-decomposition-rg}
\end{equation}
The number-conserving current has no fast mass phase, so its total
covariant derivative carries only residual momentum.  The pair currents
instead acquire a factor $\mp2im_cv_\alpha$ when differentiated.
Products with a net pair-creation or pair-annihilation number do not
contribute to the forward matrix element in the sector with one charm
and one anticharm.  The leading connected pair contribution is therefore
the product of a pair-creation current and a pair-annihilation current.

After eliminating the small heavy-field components at tree level, the pair component of the current is expanded through first order in
the residual momentum as
\begin{equation}
 \begin{aligned}
 \left.\bar c_j\Gamma_\chi^\mu c^i\right|_{c\bar c}
 &=e^{-2im_cv\cdot x}
 \left[(a_\chi^{[0]\mu})^i{}_j
       +\frac{1}{2m_c}(a_\chi^{[1]\mu})^i{}_j\right]
 +\mathrm{h.c.}
 \end{aligned}
 \label{eq:pair-component-rg}
\end{equation}
Applying the covariant derivative directly to the pair component in
eq.~\eqref{eq:pair-component-rg} and retaining terms through one residual
derivative, the derivative can be written as
\begin{equation}
\begin{aligned}
 \left.\mathcal D_\mu(\bar c_j\Gamma_\chi^\nu c^i)\right|_{c\bar c}
 &=e^{-2im_cv\cdot x}\Bigl[
 -2im_cv_\mu(a_\chi^{[0]\nu})^i{}_j
 +(\mathcal D_\mu a_\chi^{[0]\nu})^i{}_j
 -iv_\mu(a_\chi^{[1]\nu})^i{}_j\Bigr]
 +\mathrm{h.c.}
\end{aligned}
 \label{eq:pair-phase-derivative-rg}
\end{equation}
For the heavy-pair contribution, separating the fast-phase and
residual-derivative terms gives
\begin{equation}\label{eq:dimension-eight-to-seven-rg}
\begin{aligned}
 \mathcal Q[\chi,\chi']
 &=(2m_c)^2\mathcal O^{(6)}[\chi,\chi']
  +2m_c\mathcal O^{(7)}[\chi,\chi']+\cdots,
 \\
 \widetilde{\mathcal Q}[\chi,\chi']
 &=(2m_c)^2\widetilde{\mathcal O}^{(6)}[\chi,\chi']
  +2m_c\widetilde{\mathcal O}^{(7)}[\chi,\chi']+\cdots.
 \end{aligned}
 \end{equation}
Here the ellipses denote terms with two or more residual derivatives
and the associated gluonic structures; the displayed reduction is
truncated through canonical dimension seven.

With $\Trc$ a trace over color indices, the dimension-six operators and
the dimension-seven structures are
\begin{align}
 \mathcal O^{(6)}[\chi,\chi']
 &=\Trc[(a_{\chi\perp\rho}^{[0]})^\dagger]\,
   \Trc(a_{\chi'\perp}^{[0]\rho})
  +\Trc[(a_{\chi'\perp\rho}^{[0]})^\dagger]\,
   \Trc(a_{\chi\perp}^{[0]\rho}),
 \label{eq:pair-O6-rg}\\[0.2em]
 \widetilde{\mathcal O}^{(6)}[\chi,\chi']
 &=\Trc[(a_{\chi\perp\rho}^{[0]})^\dagger
         a_{\chi'\perp}^{[0]\rho}]
  +\Trc[(a_{\chi'\perp\rho}^{[0]})^\dagger
         a_{\chi\perp}^{[0]\rho}],
 \label{eq:pair-O6-tilde-rg}\\[0.7em]
 \mathcal O^{(7)}[\chi,\chi']
 &=\Trc[(a_{\chi\perp\rho}^{[1]})^\dagger]
  \Trc(a_{\chi'\perp}^{[0]\rho})
 +\Trc[(a_{\chi\perp\rho}^{[0]})^\dagger]
  \Trc(a_{\chi'\perp}^{[1]\rho})
 \nonumber\\
 &\quad +i\Trc[(a_{\chi\perp\rho}^{[0]})^\dagger]
   \Trc(v\!\cdot\mathcal D\,a_{\chi'\perp}^{[0]\rho})
 -i\Trc[v\!\cdot\mathcal D
   (a_{\chi\perp\rho}^{[0]})^\dagger]
   \Trc(a_{\chi'\perp}^{[0]\rho})
 \nonumber\\
 &\quad +i\Trc[\mathcal D_\perp^\rho
   (a_{\chi\perp\rho}^{[0]})^\dagger]
   \Trc(v\!\cdot a_{\chi'}^{[0]})
 -i\Trc[v\!\cdot(a_\chi^{[0]})^\dagger]
   \Trc(\mathcal D_{\perp\rho}a_{\chi'\perp}^{[0]\rho})
 +\mathrm{h.c.},
 \label{eq:pair-O7-rg}\\[0.5em]
 \widetilde{\mathcal O}^{(7)}[\chi,\chi']
 &=\Trc[(a_{\chi\perp\rho}^{[1]})^\dagger
        a_{\chi'\perp}^{[0]\rho}
       +(a_{\chi\perp\rho}^{[0]})^\dagger
        a_{\chi'\perp}^{[1]\rho}]
 \nonumber\\
 &\quad +i\Trc[(a_{\chi\perp\rho}^{[0]})^\dagger
   (v\!\cdot\mathcal D\,a_{\chi'\perp}^{[0]\rho})]
 -i\Trc[(v\!\cdot\mathcal D
   (a_{\chi\perp\rho}^{[0]})^\dagger)
   a_{\chi'\perp}^{[0]\rho}]
 \nonumber\\
 &\quad +i\Trc[(\mathcal D_\perp^\rho
   (a_{\chi\perp\rho}^{[0]})^\dagger)
   (v\!\cdot a_{\chi'}^{[0]})]
 -i\Trc[(v\!\cdot(a_\chi^{[0]})^\dagger)
   (\mathcal D_{\perp\rho}a_{\chi'\perp}^{[0]\rho})]
 +\mathrm{h.c.}.
 \label{eq:pair-O7-tilde-rg}
\end{align}
In eqs.~\eqref{eq:pair-O7-rg} and \eqref{eq:pair-O7-tilde-rg},
$\mathrm{h.c.}$ applies to the entire preceding right-hand side.  It supplies
the reversed chirality ordering and makes each operator Hermitian and symmetric
under interchange of $\chi$ and $\chi'$.
\Needspace{5\baselineskip}
With $\gamma_\perp^\mu=(g^{\mu\nu}-v^\mu v^\nu)\gamma_\nu$, the heavy-field
projectors give
\begin{equation}
P_-\gamma_\perp^\mu\gamma_5P_+=0,
\qquad
a_{L\perp}^{[0]\mu}=a_{R\perp}^{[0]\mu}.
\label{eq:leading-pair-vector-current-rg}
\end{equation}
The leading transverse pair current therefore has the vector spin
structure appropriate to the ${}^3S_1$ $J/\psi$.  Equations
\eqref{eq:pair-phase-derivative-rg} and
\eqref{eq:dimension-eight-to-seven-rg} identify the leading contribution
to the connected heavy-pair matrix elements of $\mathcal Q$ and
$\widetilde{\mathcal Q}$ as pair annihilation followed by pair creation.
It carries two powers of the hard mass from the fast phases, whereas
terms involving only number-conserving currents carry residual derivatives.
The operators $\mathcal O^{(7)}$ describe the first residual-momentum
correction to this leading annihilation structure.

The effective-field expansion also identifies the intermediate state to
which vacuum saturation applies.  Let
$A_\chi^\rho=\Trc(a_{\chi\perp}^{[0]\rho})$.
The untilded dimension-six operator is
$\mathcal O^{(6)}[\chi,\chi']
=A_{\chi\rho}^\dagger A_{\chi'}^\rho
 +A_{\chi'\rho}^\dagger A_\chi^\rho$.
Inserting a complete set of light states between the pair currents gives
\begin{equation}
\begin{aligned}
\langle J/\psi|A_{\chi\rho}^\dagger A_{\chi'}^\rho|J/\psi\rangle
&=\sum_X
\langle J/\psi|A_{\chi\rho}^\dagger|X\rangle
\langle X|A_{\chi'}^\rho|J/\psi\rangle\\
&\simeq
\langle J/\psi|A_{\chi\rho}^\dagger|0\rangle
\langle0|A_{\chi'}^\rho|J/\psi\rangle.
\end{aligned}
\label{eq:hqeft-vacuum-intermediate-state-rg}
\end{equation}
For the leading color-singlet $c\bar c$ component, $A_\chi^\rho$
annihilates the heavy pair and leaves the vacuum.  Nonvacuum states in
this sum arise from higher Fock components containing additional gluons
or light partons.  Retaining the dominant color-singlet component therefore
motivates the vacuum-saturation approximation~\cite{BBL:1995}.
The exchanged-color operator also contains an independent color-octet
contraction; its vacuum-to-meson current matrix element vanishes by color
conservation, and its physical contribution is omitted in this leading
color-singlet approximation.

At leading order in the hard decay kernels, no gluon is exchanged between
the heavy annihilation current and the light final-state current, so the
binding effects are contained in the physical decay constant. Independent
color-octet inputs, nonvacuum intermediate states and hard radiative
corrections lie beyond this approximation.
Hence we will  evaluate $\mathcal Q$ and $\widetilde{\mathcal Q}$ directly by
vacuum saturation, retaining the physical pair momentum $P^\mu$ and fixing
the normalization by the full-QCD decay constant.

\section{Hadronic matrix elements}
\label{sec:hadronic-matrix-elements-rg}

We define the reduced forward matrix elements by
\begin{equation}
\begin{aligned}
\mathcal M_{\mathcal Q}[\chi,\chi']
&\equiv\frac{1}{2M_{J/\psi}}
\langle J/\psi(P,\lambda)|\mathcal Q[\chi,\chi']|J/\psi(P,\lambda)\rangle,\\
\widetilde{\mathcal M}_{\mathcal Q}[\chi,\chi']
&\equiv\frac{1}{2M_{J/\psi}}
\langle J/\psi(P,\lambda)|\widetilde{\mathcal Q}[\chi,\chi']|J/\psi(P,\lambda)\rangle.
\end{aligned}
\label{eq:note-normalization}
\end{equation}
For any other local operator $\mathcal O$, we use the same normalization:
\begin{equation}
\mathcal M_{J/\psi}[\mathcal O]
=\frac{\langle J/\psi|\mathcal O|J/\psi\rangle}{2M_{J/\psi}}.
\end{equation}

Let $J_\chi^\mu=\bar c_i\Gamma_\chi^\mu c^i$ and
$J_\chi^{\mu\,i}{}_{j}=\bar c_j\Gamma_\chi^\mu c^i$.
Parity forbids the axial-vector current from creating the
$J^{PC}=1^{--}$ state from the vacuum.  Consequently, the left- and
right-handed currents have the same vacuum-to-meson matrix element,
fixed by the physical vector decay constant:
\begin{equation}
\begin{gathered}
\langle0|J_L^\mu(0)|J/\psi(P,\lambda)\rangle
=\langle0|J_R^\mu(0)|J/\psi(P,\lambda)\rangle
=M_{J/\psi}f_\psi\epsilon_\lambda^\mu,\\
P\!\cdot\epsilon_\lambda=0,\qquad
\epsilon_\lambda^*\!\cdot\epsilon_\lambda=-1.
\end{gathered}
\label{eq:note-current}
\end{equation}
In the color-singlet approximation, the open-color matrix element
is $\delta^i{}_{j}/N_c$ times that in eq.~\eqref{eq:note-current}.

Translational covariance supplies the derivatives in the full-QCD
operators without a heavy-field expansion:
\begin{equation}
\begin{aligned}
\langle0|J_\chi^\mu(x)|J/\psi(P,\lambda)\rangle
&=M_{J/\psi}f_\psi\epsilon_\lambda^\mu e^{-iP\cdot x},\\
\langle0|\partial_\alpha J_\chi^\mu(0)|J/\psi(P,\lambda)\rangle
&=-iP_\alpha M_{J/\psi}f_\psi\epsilon_\lambda^\mu,\\
\langle J/\psi(P,\lambda)|\partial_\alpha J_\chi^\mu(0)|0\rangle
&=+iP_\alpha M_{J/\psi}f_\psi^*\epsilon_\lambda^{*\mu}.
\end{aligned}
\label{eq:note-derivatives}
\end{equation}
For the open-color current, the covariant derivative has the same
matrix element with the factor $\delta^i{}_{j}/N_c$ at the order retained.
Equations~\eqref{eq:note-current} and \eqref{eq:note-derivatives} give
\begin{equation}
\langle0|\partial_\mu J_\chi^\mu|J/\psi\rangle=0,
\qquad
\langle0|\mathcal D_\mu J_\chi^{\mu\,i}{}_{j}|J/\psi\rangle=0.
\label{eq:note-divergence}
\end{equation}
Thus the divergence products in both $\mathcal Q$ and
$\widetilde{\mathcal Q}$ vanish after vacuum saturation.

For the connected annihilation part of the normal-ordered four-charm
products, vacuum saturation retains both assignments of the pair-creation
and pair-annihilation currents.  Suppressing the external-state labels,
\begin{equation}
\begin{aligned}
\langle J/\psi|\mathcal Q[\chi,\chi']|J/\psi\rangle_{\rm VS}
&\simeq
\langle J/\psi|\partial_\alpha J_{\chi\rho}|0\rangle
\langle0|\partial^\alpha J_{\chi'}^\rho|J/\psi\rangle
+(\chi\leftrightarrow\chi')\\
&=-2M_{J/\psi}^4|f_\psi|^2.
\end{aligned}
\label{eq:direct-Q-vacuum-saturation-rg}
\end{equation}
Each assignment contributes $-M_{J/\psi}^4|f_\psi|^2$, using
$P^2=M_{J/\psi}^2$ and
$\epsilon_\lambda^*\!\cdot\epsilon_\lambda=-1$.
For the exchanged color contraction,
$\sum_{i,j}\delta^i{}_{j}\delta^j{}_{i}/N_c^2=1/N_c$, giving, for every chirality pair,
\begin{equation}
\mathcal M_{\mathcal Q}[\chi,\chi']
\simeq-M_{J/\psi}^3|f_\psi|^2,
\qquad
\widetilde{\mathcal M}_{\mathcal Q}[\chi,\chi']
\simeq-\frac{M_{J/\psi}^3|f_\psi|^2}{N_c}.
\label{eq:note-results}
\end{equation}
No separate residual-energy matrix element is needed here.
Inserting eq.~\eqref{eq:note-results} into eqs.~\eqref{eq:T8-W-rg}--\eqref{eq:T8-WZ-rg} gives
\begin{align}
\Gamma_{\rm WE}
&=\frac{G_F^2M_{J/\psi}^3|f_\psi|^2}{12\pi N_c}
\sum_{q_1,q_2}|V_{cq_1}|^2|V_{cq_2}|^2(C_1+N_cC_2)^2,
\label{eq:direct-W-width-rg}\\
\Gamma_{{\rm NC},q}
&=\frac{G_F^2M_{J/\psi}^3|f_\psi|^2}{3\pi}
\sum_{q,\chi,\chi',\chi''}(g_{Z,\chi''}^{q})^2
 g_{Z,\chi}^{c}g_{Z,\chi'}^{c}
\left[\mathcal K_{ZZ}^{\chi\chi';\chi''}
+\frac{\widetilde{\mathcal K}_{ZZ}^{\chi\chi';\chi''}}{N_c}\right],
\label{eq:direct-Z-quark-width-rg}\\
\Gamma_{{\rm NC},\ell}
&=\frac{G_F^2M_{J/\psi}^3|f_\psi|^2}{3\pi}
\sum_{f,\chi,\chi',\chi''}(g_{Z,\chi''}^{f})^2
 g_{Z,\chi}^{c}g_{Z,\chi'}^{c},
\label{eq:direct-Z-lepton-width-rg}\\
\Gamma_{\rm int}
&=\frac{G_F^2M_{J/\psi}^3|f_\psi|^2}{3\pi}
\sum_{q_1,\chi}|V_{cq_1}|^2g_{Z,L}^{q_1}g_{Z,\chi}^{c}
\left[\mathcal K_{WZ}^{\chi}
+\frac{\widetilde{\mathcal K}_{WZ}^{\chi}}{N_c}\right].
\label{eq:direct-WZ-width-rg}
\end{align}
The subscripts denote the same shorthand as in the transition operators.

\section{Numerical analysis}
\label{sec:numerical-analysis-rg}

We combine the short-distance coefficients with the matrix elements of
section~\ref{sec:hadronic-matrix-elements-rg} to evaluate the inclusive weak
width at the central renormalization scale $\mu_0=M_{J/\psi}$.
We assign a scale uncertainty from the envelope obtained by varying a
common scale over $1~\mathrm{GeV}\leq\mu\leq3M_{J/\psi}$, with all
masses and hadronic matrix elements held fixed. The upper and lower errors
are the maximal upward and downward shifts from the central value over
the full interval. For each total width or branching fraction, all
contributions are evaluated at the same scale before taking the envelope.
We quote two errors: the perturbative scale uncertainty and an assigned
hadronic HQE uncertainty. The first estimates missing higher-order
perturbative corrections; the hard kernels remain at tree level.
The second combines vacuum saturation and the kinetic/chromomagnetic
inputs through their joint variation at $\mu_0$.

Table~\ref{tab:numerical-inputs-rg} summarizes the common inputs.
The single-charm kernels and matrix-element estimates use $m_c=1.48~\mathrm{GeV}$~\cite{King:2021}.
The directly saturated annihilation matrix elements instead use the
physical pair mass $M_{J/\psi}$, as in eq.~\eqref{eq:note-results}.
The value $m_b=4.186~\mathrm{GeV}$ in
table~\ref{tab:numerical-inputs-rg} serves only as the flavor threshold in
the leading-logarithmic evolution of the Wilson coefficients, which uses
one-loop running.
The masses, widths, CKM elements, and electroweak inputs are taken
from the Review of Particle Physics~\cite{PDG:2026}, with the CKM
elements from its global fit and
$s_W^2=\sin^2\widehat\theta_W(M_Z)$ in the $\overline{\mathrm{MS}}$
scheme.
The decay constant follows from the leptonic width,
$\Gamma(J/\psi\to e^+e^-)=4\pi\alpha^2Q_c^2f_\psi^2/(3M_{J/\psi})$, with
$\Gamma(J/\psi\to e^+e^-)=5.53~\mathrm{keV}$~\cite{PDG:2026}, $Q_c=2/3$, and
the effective coupling $\alpha(M_{J/\psi}^2)=1/134.02$~\cite{HPQCD:2020fpsi},
which gives $f_\psi=0.4065~\mathrm{GeV}$.

\begin{table}[!htbp]
 \centering
 \small
 \renewcommand{\arraystretch}{1.15}
 \begin{tabular}{@{}lclc@{}}
 \toprule
 Parameter & Value & Parameter & Value \\
 \midrule
 $G_F$ & $1.1663785\times10^{-5}~\mathrm{GeV}^{-2}$
 & $N_c$ & $3$ \\
 $m_c$ & $1.48~\mathrm{GeV}$
 & $M_{J/\psi}$ & $3.096900~\mathrm{GeV}$ \\
 $M_{\eta_c}$ & $2.98409~\mathrm{GeV}$
 & $\Gamma_{J/\psi}^{\rm tot}$ & $92.6~\mathrm{keV}$ \\
 $\lvert V_{cd}\rvert$ & $0.22503$
 & $\lvert V_{cs}\rvert$ & $0.97345$ \\
 $\lvert V_{ud}\rvert$ & $0.97431$
 & $\lvert V_{us}\rvert$ & $0.22517$ \\
 $s_W^2$ & $0.23122$
 & $\alpha_s(M_Z)$ & $0.1180$ \\
 $M_Z$ & $91.1879~\mathrm{GeV}$
 & $M_W$ & $80.3625~\mathrm{GeV}$ \\
  $m_b$ & $4.186~\mathrm{GeV}$
 & $f_\psi$ & $0.4065~\mathrm{GeV}$ \\
 \bottomrule
 \end{tabular}
 \caption{Numerical inputs held fixed throughout the scale variation.}
 \label{tab:numerical-inputs-rg}
\end{table}
\Needspace{9\baselineskip}
Across the bottom threshold, the leading-logarithmic evolution is
\begin{equation}
\begin{aligned}
 U_{\gamma_0}(\mu_H,\mu)
 &=\begin{cases}
 \displaystyle\left[\frac{\alpha_s^{(5)}(\mu_H)}{\alpha_s^{(5)}(\mu)}\right]^%
 {\gamma_0/(2\beta_0^{(5)})}, & \mu\geq m_b,\\[1.2em]
 \displaystyle\left[\frac{\alpha_s^{(5)}(\mu_H)}{\alpha_s^{(5)}(m_b)}\right]^%
 {\gamma_0/(2\beta_0^{(5)})}
 \left[\frac{\alpha_s^{(4)}(m_b)}{\alpha_s^{(4)}(\mu)}\right]^%
 {\gamma_0/(2\beta_0^{(4)})}, & \mu<m_b,
 \end{cases}\\[0.5em]
 \beta_0^{(n_f)}&=11-\frac{2n_f}{3} .
\end{aligned}
 \label{eq:scale-evolution-factor-rg}
\end{equation}
The coupling is continuous at $m_b$ at this order. The charged-current
coefficients are $C_1\pm C_2=U_{\gamma_0}(M_W,\mu)$ with $\gamma_0=4,-8$,
and the neutral-current coefficients are $C_i^Z=U_{\gamma_0}(M_Z,\mu)$ with
$\gamma_0=4,-8,-16,2$ for $i=1,\dots,4$~\cite{Du:2024nc}.
Table~\ref{tab:scale-coefficients-rg} lists the running coupling,
the charged-current and neutral-current Wilson coefficients, and the
derived single-charm coefficients $c_3,c_5$, with their central values
at $\mu_0=M_{J/\psi}$ and their values at the two endpoints,
$\mu=1~\mathrm{GeV}$ and $\mu=3M_{J/\psi}$.
\begin{table}[!htbp]
 \centering
 \small
 \renewcommand{\arraystretch}{1.15}
 \setlength{\tabcolsep}{10pt}
 \begin{tabular}{@{}l*{3}{D{.}{.}{-1}}@{}}
 \toprule
 Quantity & \multicolumn{1}{c}{$\mu=1~\mathrm{GeV}$}
 & \multicolumn{1}{c}{$\mu=M_{J/\psi}$}
 & \multicolumn{1}{c}{$\mu=3M_{J/\psi}$} \\
 \midrule
 $\alpha_s(\mu)$ & 0.3551 & 0.2317 & 0.1758 \\
 $C_1$ & 1.242 & 1.124 & 1.063 \\
 $C_2$ & -0.4802 & -0.2796 & -0.1570 \\
 $C_1^Z$ & 0.7583 & 0.8401 & 0.9012 \\
 $C_2^Z$ & 1.739 & 1.417 & 1.231 \\
 $C_3^Z$ & 3.024 & 2.007 & 1.516 \\
 $C_4^Z$ & 0.8708 & 0.9166 & 0.9493 \\
 $c_3$ & 6.117 & 5.385 & 5.118 \\
 $c_5$ & -2.707 & -5.752 & -7.572 \\
 \bottomrule
 \end{tabular}
 \caption{Running coupling and short-distance coefficients at three
 renormalization scales. The middle column corresponds to the central
 scale $\mu_0=M_{J/\psi}$.}
 \label{tab:scale-coefficients-rg}
\end{table}

For the single-charm sector, we estimate the kinetic matrix element
from the spin-averaged $1S$ mass, in which the leading spin--spin
interaction cancels, and the chromomagnetic matrix element from the
hyperfine splitting \cite{Beneke:1996}. We adopt estimated relative
hadronic uncertainties of order
$\varepsilon_{\rm had}=\Lambda_{\rm QCD}/m_c\simeq0.20$ for vacuum
saturation and for each of these two inputs:
\begin{equation}
 \begin{aligned}
 \mu_\pi^2&=(0.161\pm0.032)~\mathrm{GeV}^2,\\
 \mu_G^2&=(-0.0835\pm0.0167)~\mathrm{GeV}^2.
 \end{aligned}
 \label{eq:numerical-nonperturbative-parameters-rg}
\end{equation}
The central kinetic estimate follows from
$\tfrac14(3M_{J/\psi}+M_{\eta_c})=2m_c+\mu_\pi^2/m_c+\langle V\rangle$
with the spin-independent potential energy $\langle V\rangle$ neglected,
and corresponds to $\langle v^2\rangle=\mu_\pi^2/m_c^2\simeq0.073$.
The central hyperfine estimate,
$\mu_G^2=-m_c(M_{J/\psi}-M_{\eta_c})/2$, assumes a tree-level
spin--spin interaction and a common spatial wave function for
$J/\psi$ and $\eta_c$.
The scalar and chromomagnetic matrix elements are
\cite{King:2021,Beneke:1996}
\begin{equation}
 \begin{aligned}
 \mathcal M_{J/\psi}[\mathcal O_3]
 &=2\left(1-\frac{\mu_\pi^2}{2m_c^2}
            +\frac{\mu_G^2}{2m_c^2}\right),
 &\mathcal M_{J/\psi}[\mathcal O_5]&=2\mu_G^2 .
 \end{aligned}
 \label{eq:single-charm-matrix-elements-rg}
\end{equation}
They enter the sum of the charm and anticharm decay widths as
\begin{equation}
 \Gamma_{c+\bar c}
 =\frac{G_F^2m_c^5}{192\pi^3}
 \left[c_3\mathcal M_{J/\psi}[\mathcal O_3]
       +\frac{c_5}{m_c^2}\mathcal M_{J/\psi}[\mathcal O_5]\right].
 \label{eq:single-charm-width-rg}
\end{equation}

For the hadronic HQE uncertainty, we multiply all saturated $\mathcal M_{\mathcal Q}$
and $\widetilde{\mathcal M}_{\mathcal Q}$ by one common factor
$1+\delta_{\rm VS}$, with $|\delta_{\rm VS}|\leq\varepsilon_{\rm had}$.
Together with $\delta_{\rm VS}$, we vary $\mu_\pi^2$ and $\mu_G^2$ independently by
$\varepsilon_{\rm had}$ of their magnitudes, keeping the same $\mu_G^2$ in both
matrix elements. At $\mu_0$, the envelope over this three-input box gives
the single HQE error quoted below. These assigned theory ranges are not
statistical errors and are not combined in quadrature.
For a common normalization, table~\ref{tab:matrix-elements-rg} lists
dimensionless combinations obtained by dividing each reduced matrix element
by the appropriate power of $m_c$. The corresponding unnormalized matrix
elements enter the width formulas. The
heavy-pair entries apply to every chirality pair $\chi,\chi'$ at the order
considered.
\begin{table}[!htbp]
 \centering
 \small
 \renewcommand{\arraystretch}{1.15}
 \setlength{\tabcolsep}{10pt}
 \begin{tabular}{@{}lc@{}}
 \toprule
 Dimensionless combination & Central value with hadronic uncertainty \\
 \midrule
 $\mathcal M_{J/\psi}[\mathcal O_3]$
 & $(1.89\pm0.02)$ \\
 $\mathcal M_{J/\psi}[\mathcal O_5]/m_c^2$
 & $(-0.0762\pm0.0152)$ \\
 $\mathcal M_{\mathcal Q}[\chi,\chi']/m_c^5$
 & $(-0.691\pm0.138)$ \\
 $\widetilde{\mathcal M}_{\mathcal Q}[\chi,\chi']/m_c^5$
 & $(-0.230\pm0.046)$ \\
 \bottomrule
 \end{tabular}
 \caption{Dimensionless reduced matrix elements with assigned hadronic HQE
 uncertainties. The four-charm entries share a common vacuum-saturation
 normalization factor, and the single-charm entries share the same kinetic
 and chromomagnetic inputs.}
 \label{tab:matrix-elements-rg}
\end{table}

\FloatBarrier
At each scale, the weak width is organized into the single-charm sector
and two heavy-pair annihilation sectors as
\begin{equation}
 \Gamma_{\rm weak}
 =\Gamma_{\lvert\Delta C\rvert=1}
 +\Gamma_{\Delta C=0,\,\lvert\Delta S\rvert = 1 }
 +\Gamma_{\Delta C=0,\,\Delta S=0} .
 \label{eq:flavor-width-decomposition-rg}
\end{equation}
The first term equals $\Gamma_{c+\bar c}$ of
eq.~\eqref{eq:single-charm-width-rg} and includes both $c$ and $\bar c$
decays. In the $W$-exchange contribution, diagonal
flavor terms belong to $\Delta S=0$, while off-diagonal terms have
$\lvert\Delta S\rvert=1$ at the order considered. The neutral-current
and $W$--$Z$ interference terms contribute only to $\Delta S=0$.
Table~\ref{tab:flavor-classified-widths-rg} lists $\Gamma_{\rm WE}$,
$\Gamma_{\rm NC}$, and $\Gamma_{\rm int}$ from
eqs.~\eqref{eq:direct-W-width-rg}--\eqref{eq:direct-WZ-width-rg} by flavor
sector.
\begin{table}[!htbp]
 \centering
 \small
 \renewcommand{\arraystretch}{1.15}
 \begin{tabular*}{\textwidth}{@{\extracolsep{\fill}}lccc@{}}
 \toprule
 Contribution
 & $\lvert\Delta C\rvert=1$
 & $\Delta C=0,\ \lvert\Delta S\rvert=1$
 & $\Delta C=0,\ \Delta S=0$ \\
 \midrule
 $\Gamma_{c+\bar c}$
 & $1.72^{+0.19}_{-0.06}\pm0.02$ & -- & -- \\
 $\Gamma_{\rm WE}$
 & -- & $0.0460^{+0.1522}_{-0.0460}\pm0.0092$
 & $0.432^{+1.429}_{-0.432}\pm0.086$ \\
 $\Gamma_{\rm NC}$
 & -- & -- & $6.85^{+0.34}_{-0.15}\pm1.37$ \\
 $\Gamma_{\rm int}$
 & -- & -- & $-1.68^{+2.89}_{-1.75}\pm0.34$ \\
 \midrule
 Total & $1.72^{+0.19}_{-0.06}\pm0.02$
 & $0.0460^{+0.1522}_{-0.0460}\pm0.0092$
 & $5.60^{+3.01}_{-0.47}\pm1.12$ \\
 \bottomrule
 \end{tabular*}
 \caption{Signed contributions in units of $10^{-12}~\mathrm{GeV}$,
 classified by flavor change. Central values use $\mu_0=M_{J/\psi}$;
 the first errors are scale envelopes over
 $1~\mathrm{GeV}\leq\mu\leq3M_{J/\psi}$, followed by the combined
 hadronic HQE errors. Contributions share the underlying input variations,
 and each total is varied as a whole.}
 \label{tab:flavor-classified-widths-rg}
\end{table}

The flavor decomposition in table~\ref{tab:flavor-classified-widths-rg}
shows that the $\Delta C=0,\Delta S=0$ sector dominates throughout the scale interval,
with the neutral-current term providing the largest contribution within that
sector.  In the order of the three sectors in
eq.~\eqref{eq:flavor-width-decomposition-rg}, the fractions of the weak
width at the central scale are $(23.4\%,\,0.62\%,\,76.0\%)$, so pair
annihilation provides $76.6\%$.
The single-charm width increases moderately as the scale is lowered,
through the changes in $c_3,c_5$.
For $W$ exchange, color-singlet factorization makes the width
proportional to $(C_1+N_cC_2)^2$; the smaller
$\lvert C_1+N_cC_2\rvert$ at $1~\mathrm{GeV}$ therefore suppresses
both flavor sectors relative to the central values. This combination
vanishes near $\mu=1.48~\mathrm{GeV}$, so the lower scale envelopes
of both $W$-exchange contributions reach zero; their upper envelopes
occur at $\mu=3M_{J/\psi}$. The light-quark
component drives the scale variation of the neutral-current contribution,
while its leptonic component remains unchanged.
The $W$--$Z$ interference depends linearly on $C_1+N_cC_2$.
It is destructive at the central scale $\mu_0$ and becomes constructive
below the zero of this combination. This sign change within the scale
interval drives the large upward scale uncertainty.

\FloatBarrier
\begin{table}[!htbp]
 \centering
 \small
 \renewcommand{\arraystretch}{1.15}
 \setlength{\tabcolsep}{10pt}
 \begin{tabular}{@{}lc@{}}
 \toprule
 Quantity & Central value with separate uncertainties \\
 \midrule
 $\Gamma_{\Delta C=0}~[\mathrm{GeV}]$
 & $(5.65^{+2.99}_{-0.35}\pm1.13)\times10^{-12}$ \\
 $\Gamma_{\rm weak}~[\mathrm{GeV}]$
 & $(7.37^{+3.18}_{-0.40}\pm1.15)\times10^{-12}$ \\
 $\mathcal B_{\Delta C=0}$
 & $(6.10^{+3.23}_{-0.38}\pm1.22)\times10^{-8}$ \\
 $\mathcal B_{\rm weak}$
 & $(7.96^{+3.43}_{-0.43}\pm1.24)\times10^{-8}$ \\
 $\mathcal B_{\rm WE}$
 & $(5.16^{+17.07}_{-5.16}\pm1.03)\times10^{-9}$ \\
 \bottomrule
 \end{tabular}
 \caption{Pair-annihilation and inclusive weak widths and branching
 fractions, and the pure $W$-exchange branching fraction, at
 $\mu_0=M_{J/\psi}$. Errors are quoted as (scale, hadronic HQE).}
 \label{tab:total-widths-branching-rg}
\end{table}

We denote by $\Gamma_{\Delta C=0}$ the total pair-annihilation width,
i.e.\ the sum of the two $\Delta C=0$ terms in
eq.~\eqref{eq:flavor-width-decomposition-rg}, including the $W$-exchange, neutral-current and $W$--$Z$ interference
contributions.
Branching fractions are defined as
$\mathcal B_X=\Gamma_X/\Gamma_{J/\psi}^{\rm tot}$, using the total width
in table~\ref{tab:numerical-inputs-rg}.
We also quote the pure $W$-exchange annihilation contribution
$\mathcal B_{\rm WE}=\Gamma_{\rm WE}/\Gamma_{J/\psi}^{\rm tot}$, with $\Gamma_{\rm WE}$
summed over both strangeness sectors in
table~\ref{tab:flavor-classified-widths-rg}.
Adding the $|\Delta C|=1$ single-charm decays gives the branching
fraction of the contribution mediated by the $W$ boson alone,
$(\Gamma_{\rm WE}+\Gamma_{|\Delta C|=1})/\Gamma_{J/\psi}^{\rm tot}
=(2.37^{+1.64}_{-0.42}\pm0.13)\times10^{-8}$.
Table~\ref{tab:total-widths-branching-rg} summarizes the results.
For comparison, the measured $D$-meson lifetimes give the central widths~\cite{PDG:2026}
\begin{equation}
(\Gamma_{D^0},\Gamma_{D^+},\Gamma_{D_s^+})
=(1.60,0.637,1.31)\times10^{-12}~\mathrm{GeV}.
\label{eq:D-meson-widths-rg}
\end{equation}
At the central scale,
$\Gamma_{\rm WE}=4.78\times10^{-13}~\mathrm{GeV}$ and
$\Gamma_{|\Delta C|=1}=1.72\times10^{-12}~\mathrm{GeV}$, giving
$\Gamma_{\rm WE}+\Gamma_{|\Delta C|=1}=2.20\times10^{-12}~\mathrm{GeV}$.
Using unrounded central inputs, the ratios of this sum to the widths in
eq.~\eqref{eq:D-meson-widths-rg} are $1.37$, $3.45$, and
$1.67$, respectively.

The present results omit radiative corrections to the hard
kernels, hadronic effects beyond vacuum saturation, single-charm terms of
dimension six and higher such as the Darwin
contribution~\cite{Lenz:2020Darwin,King:2021}, heavy-pair operators with
independent relative-motion or gluonic structure, penguin and
flavor-singlet mixing in the Wilson-coefficient
evolution~\cite{Jenkins:2017jig}, and final-state masses. Uncertainties
from the decay constant and the charm-mass prescription are not
propagated.

\FloatBarrier
\section{Conclusions}
\label{sec:conclusions}

We have calculated the inclusive weak width of the $J/\psi$ through the
optical theorem, including single-charm decays, charged- and
neutral-current pair annihilation, and their interference. The
heavy-quark effective-field expansion identifies the leading
pair-annihilation structure and motivates vacuum saturation, and the
full-QCD matrix elements are fixed by the physical $J/\psi$ mass and
decay constant.
With tree-level hard kernels and leading-logarithmic Wilson coefficients
at $\mu_0=M_{J/\psi}$, we obtain
$\mathcal B_{\rm weak}=(7.96^{+3.43}_{-0.43}\pm1.24)\times10^{-8}$,
where the errors are the scale envelope over
$1~\mathrm{GeV}\leq\mu\leq3M_{J/\psi}$ and the hadronic HQE
uncertainty. The contribution mediated by the $W$ boson alone is
$(2.37^{+1.64}_{-0.42}\pm0.13)\times10^{-8}$, and the
$(\Delta C,|\Delta S|)=(0,1)$ and $|\Delta C|=1$ sectors give
$(4.97^{+16.44}_{-4.97}\pm0.99)\times10^{-10}$ and
$(1.86^{+0.20}_{-0.06}\pm0.02)\times10^{-8}$, respectively.

Pair annihilation dominates, providing about $77\%$ of the central
width, with the neutral-current term as the largest contribution. The
$W$--$Z$ interference changes sign within the scale interval and
drives the large upward scale uncertainty. Estimates based on single-charm
decay alone therefore miss most of the $J/\psi$ weak width. Quantitative
refinement requires radiative corrections to the hard kernels, consistent
evolution of the hadronic matrix elements, and hadronic inputs beyond
vacuum saturation.

\acknowledgments

This work was supported in part by the National Natural Science Foundation
of China under Grant No.~12575096.

\FloatBarrier
\begingroup
\raggedright

\endgroup

\end{document}